\documentclass[twocolumn,nofootinbib]{revtex4}
\usepackage{graphicx}
\usepackage{latexsym}

\def\be{\begin{equation}}
\def\ee{\end{equation}}
\def\bea{\begin{eqnarray}}
\def\eea{\end{eqnarray}}

\begin{document}

\title{Constraints on Models with a Break in the Primordial Power Spectrum}

\author{Hong Li${}^{a,b,c}$}
\author{Jun-Qing Xia${}^{d}$}
\author{Robert Brandenberger${}^{e,a,b,c}$}
\author{Xinmin Zhang${}^{a,b}$}

\affiliation{${}^a$ Institute of High Energy Physics, Chinese
Academy of Science, P.O.Box 918-4, Beijing 100049, P.R.China}

\affiliation{${}^b$ Theoretical Physics Center for Science
Facilities (TPCSF), Chinese Academy of Science, P.R.China}

\affiliation{${}^c$ Kavli Institute for Theoretical Physics, Chinese
Academy of Science, Beijing 100190, P. R. China}

\affiliation{${}^d$Scuola Internazionale Superiore di Studi
Avanzati, Via Beirut 2-4, I-34014 Trieste, Italy}

\affiliation{${}^e$ Department of Physics, McGill University, 3600
University Street, Montreal, QC, H3A 2T8, Canada}


\begin{abstract}

One of the characteristics of the ``Matter Bounce" scenario, an
alternative to cosmological inflation for producing a
scale-invariant spectrum of primordial adiabatic fluctuations on
large scales, is a break in the power spectrum at a characteristic
scale, below which the spectral index changes from $n = 1$ to $n =
3$. We study the constraints which current cosmological data place
on the location of such a break, and more generally on the position
of the break and the slope at length scales smaller than the break.
The observational data we use include the WMAP five-year data set
(WMAP5), other CMB data from BOOMERanG, CBI, VSA, and ACBAR,
large-scale structure data from the Sloan Digital Sky Survey (SDSS,
their luminous red galaxies sample), Type Ia Supernovae data
(the ``Union" compilation), and the Sloan Digital Sky Survey
Lyman-$\alpha$ forest power spectrum (Ly$\alpha$) data. We employ
the Markov Chain Monte Carlo method to constrain the features in the
primordial power spectrum which are motivated by the matter bounce
model. We give an upper limit on the length scale where the break in
the spectrum occurs.

\end{abstract}


\maketitle


\section{Introduction}
\label{Int}

Most parameter fitting studies assume that the spectrum of
primordial cosmological perturbations is a fixed power of the
co-moving wave-number $k$ over the range of scales being probed by
observations. At most, a parameter describing a running of the
spectrum is introduced. However, it is possible that the primordial
spectrum has particular features which are poorly described assuming
a slow running superimposed over a spectrum with fixed slope. In
fact, it was found \cite{Souradeep} that making an ansatz for the
spectrum which includes special features may improve the fit to the
data.

Recently, it has been realized that the ``matter bounce" provides a possible alternative
to cosmological inflation for providing a spectrum of scale-invariant adiabatic
fluctuations on large scales. As realized in \cite{Wands, FB1,Wands2} (see also
\cite{Starob}), initial perturbations starting in their vacuum state and exiting the
Hubble radius in a matter-dominated phase of contraction lead to a scale-invariant
spectrum of curvature fluctuations on super-Hubble scales in the contracting phase
\footnote{For super-Hubble scale fluctuations it is important to specify the gauge:
we are talking about curvature fluctuations in co-moving gauge.}. If the
contracting phase connects to an expanding phase via a non-singular bounce,
then, as studied in \cite{Cai2008} in the context of the Lee-Wick bounce,
a particular realization of a non-singular bounce, then the co-moving curvature
fluctuations pass through the bounce without change in the spectral slope, as
long as scales which are large compared to the duration of the bounce are
considered (see \cite{Omid,Biswas3,Cai2007} for other studies of this
transfer of fluctuations reaching the same conclusion).

In a contracting universe, the energy density of radiation increases faster than
that of matter. Hence, if the initial state contained both matter and radiation,
it is inevitable that a change in the equation of state from matter to radiation
will occur, very much like a transfer from radiation to matter occurs in the
expanding branch. As we show in the following section, scales which
exit the Hubble radius in the radiation phase acquire an $n = 3$ spectrum.
If the cosmological bounce is close to time-symmetric, we should expect
the break in the slope of the spectrum of primordial perturbations to occur
on a cosmologically interesting scale, close to the scale which re-enters
the Hubble radius during the expanding phase at the time $t_{eq}$ of
equal matter and radiation.

In this Note we study the constraints which can be put from current
observations on the
transition scale where the spectral index changes from $n = 3$ to $n
= 1$ . Comparisons of a scale dependent
primordial power spectrum with observational data can also been
found in Ref.\cite{feng}.

The outline of this Note is as follows: in the following Section we explain
why in a matter bounce model a break in the spectrum from $n = 1$ on
large scales to $n = 3$ on small scales will occur. In Section 3 we will
discuss our method and the data used. Our numerical results are
presented in Section 4, and we end with a discussion.


\section{Break in the Spectrum in the Matter Bounce Scenario}

Let us review the computation of the spectrum of cosmological perturbations
on super-Hubble scales starting from vacuum initial fluctuations on
sub-Hubble scales. We work in longitudinal gauge (see e.g. \cite{MFB}
for a detailed review of the theory of cosmological fluctuations and
\cite{RHBrev} for a shorter overview) in which the metric is given by
\be
ds^2 \, = \, a^2(\eta) \bigl[ (1 + 2 \Phi) d\eta^2 - (1 - 2 \Phi) d{\bf{x}}^2 \bigr] \, ,
\ee
where $\eta$ is conformal time, the ${\bf{x}}$ denote co-moving spatial
coordinates and $\Phi({\bf{x}}, \eta)$ is the generalized gravitational
potential which carries the information about the fluctuations. As long
as we are far from the bounce time, a good quantity to characterize the
magnitude of the inhomogeneities is $\zeta$, the curvature fluctuation
in co-moving coordinates which is expressed in terms of $\Phi$ via
\be
\zeta \, = \, \frac{2}{3} \bigl( {\cal{H}} \Phi^{'} + \Phi \bigr) \frac{1}{1 + w} + \Phi \, ,
\ee
$\cal{H}$ denoting the Hubble expansion rate in conformal time, a prime
indicating the derivative with respect of $\eta$, and $w = p / \rho$ being
the equation of state parameter ($p$ and $\rho$ are pressure and
energy density, respectively).

The variable $\zeta$ is closely related to the variable $v$
\cite{MFB} in terms of which the action for cosmological
fluctuations has canonical kinetic term:
\be \label{rel}
\zeta \, = \, \frac{v}{z} \, ,
\ee
where $z(\eta)$ is a function of the cosmological background. If matter
is modeled as a scalar field $\varphi$, then
\be
z \, = \, a \frac{\cal{H}}{\varphi^{\prime}} \, .
\ee
If the equation of state is not changing, then $z \sim a$.
Vacuum initial conditions on sub-Hubble scales hence imply that
on such scales
\be \label{vac}
v(k, t) \, = \, \frac{1}{\sqrt{2k}} \, .
\ee

The equation of motion for the Fourier mode $v_k$ of $v$ is
\be
v_k^{''} + \bigl( k^2 - \frac{z^{''}}{z} \bigr) v_k \, = \, 0 \, .
\ee
From this it follows that $v$ is oscillating on sub-Hubble scales,
whereas on super-Hubble scales the evolution of $v$ is determined by
the function $z$.  If the background scale factor evolves as 
\be
a(t) \, \sim \, t^p \, ,
\ee
then the super-Hubble solutions scale as
\be \label{scaling}
v(\eta) \, \sim \, \eta^{\alpha}
\ee
with
\be \label{alpha}
\alpha \, = \, \frac{1}{2} \pm \nu \,\,\,\, , \,\,\,\, \nu \, = \, \frac{1}{2} \frac{1 - 3p}{1 - p} \, .
\ee

In the matter phase we have $p = 2/3$ and hence $\nu = - 3/2$ which leads to the
two values for $\alpha$ which are $\alpha = 2$ and $\alpha = -1$. The second
mode is growing in the contracting phase. Making use of vacuum initial conditions
(\ref{vac}) at Hubble radius crossing, we obtain a scale-invariant power spectrum
on super-Hubble scales
\bea \label{powerm}
P_{\zeta}(k, \eta) \, &\sim& \, k^3 |\zeta_k(\eta)|^2 \, \sim \, k^3 |v_k(\eta)|^2 \nonumber \\
&\sim& k^3 |v_k(\eta_H(k))|^2 \eta_H(k)^2 \, \sim \, {\rm const} \, ,
\eea
where the first step is the definition of the power spectrum, the second step uses
(\ref{rel}), the third the temporal scaling (\ref{scaling}) of the dominant mode,
and in the final step we have inserted the vacuum initial conditions (\ref{vac}),
the Hubble crossing relation $\eta_H(k) \sim k^{-1}$ and the scaling of the scale
factor as a function of conformal time (see also \cite{Cai2008} where the
calculation was done following the evolution of $\Phi$ from initial Hubble
radius crossing in the contracting phase until the post-bounce expanding
phase).

In the radiation phase (see \cite{RX} for an earlier analysis of the evolution
of fluctuations in a contracting radiation phase)
we have $p = 1/2$ and hence $\nu = - 1/2$ which leads to
the two values for $\alpha$ which are $\alpha = 1$ and $\alpha = 0$. Hence
\be
v_k(\eta) \, = \, c_1 \eta + c_2 \, ,
\ee
where $c_1$ and $c_2$ are constants. Thus, the dominant mode of
$v$ is constant on super-Hubble scales (the corresponding mode of $\zeta$
is growing). A calculation analogous to (\ref{powerm}) yields
\bea \label{powerr}
P_{\zeta}(k, \eta) \, &\sim& \, k^3 |\zeta_k(\eta)|^2 \, \sim \, k^3 |v_k(\eta)|^2 \nonumber \\
&\sim& k^3 |v_k(\eta_H(k))|^2  \, \sim \, {\rm k^2} \, ,
\eea
where we have used the constancy of $v$ on super-Hubble scales and the
vacuum initial conditions. This is an $n_s = 3$ spectrum.

To summarize this section, we have shown that inhomogeneities originating
as quantum vacuum perturbations on sub-Hubble scales and crossing the
Hubble radius in a matter-dominated contracting phase obtain a scale-invariant
$n_s = 1$ spectrum, whereas those exiting the Hubble radius during a
radiation-dominated contracting phase get an $n_s = 3$ spectrum.

If our universe has emerged from a time-symmetric bounce, we would expect
all scales which enter the Hubble radius after the time $t_{eq}$ of equal
matter and radiation to have a scale-invariant spectrum, whereas those re-entering
earlier would have an $n_s = 3$ spectrum. If the bounce is asymmetric because
of the production of radiation around the bounce point, we would expect that
the transition scale is somewhat smaller than the scale which re-enters the
Hubble radius at $t_{eq}$. In the following, we will study what constraints which
current data can give on the scale at which the break in the index of the
power spectrum occurs.

\section{Method and Data}
\label{Method}

In our study, we perform a global analysis using the publicly
available MCMC package CosmoMC\footnote{Available at:
http://cosmologist.info/cosmomc/.} \cite{CosmoMC}. Our theory input
is as follows: We assume purely adiabatic initial conditions. Our
theory parameter space vector is: 
\begin{equation} \label{parameter}
{\bf P} \, \equiv \, (\omega_{b}, \omega_{c},
 \Theta_{s}, \tau, n_{s}, A_{s}, k_{s}, n_{s2})~,
\end{equation}
where $\omega_{b}\equiv\Omega_{b}h^{2}$ and
$\omega_{c}\equiv\Omega_{c}h^{2}$, in which $\Omega_{b}$ and
$\Omega_{c}$ are the physical baryon and cold dark matter densities
relative to the critical density,  $\Theta_{s}$ is the ratio
(multiplied by 100) of the sound horizon to the angular diameter
distance at decoupling, $\tau$ is the optical depth to
re-ionization. The remaining parameters are related to
the primordial scalar power spectrum
$\mathcal{P}_{\chi}(k)$ which is parameterized as \cite{Ps}:
\begin{equation}
  {\ln\mathcal{P}_{\chi}(k)} \, = \, \left\{ \begin{array}
 {cc} \ln A_s(k_{s0})+(n_s-1)\ln\left(\frac{k}{k_{s0}}\right), {k\leq k_{s}},\\
\ln A_s^{\prime}(k_{s0})+(n_{s2}-1)\ln\left(\frac{k}{k_{s0}}\right), {k> k_{s}}    .\\
\end{array} \right.
\label{eqn:yequation}
\end{equation}
where $A_s(k_{s0})$ and $A_s^{\prime}(k_{s0})$ are defined as the
amplitudes of initial power spectrum before and after the step,
$n_s$ and $n_{s2}$ measure the spectral index before and after the
step in the power spectrum, a step which is located at the co-moving
wavenumber $k_{s}$, and we match the power spectrum at $k_{s}$
during the numerical calculation in order to guarantee the
continuity of the power spectrum. For the pivot scale we set
$k_{s0}=0.05$Mpc$^{-1}$. We assume that the universe is spatially
flat and work in the context of the $\Lambda$CDM framework with
equation of state $w=-1$ for the dark energy component.

  In our calculations, we have
taken the total likelihood to be the products of the separate
likelihoods of CMB, SNIa, LSS and Lya. Alternatively defining
$\chi^2 = -2 \log {\bf \cal{L}}$, we get \be \chi^2_{total} =
\chi^2_{CMB}+ \chi^2_{SNIa}+\chi^2_{LSS}+\chi^2_{Lya}~~~~ .\ee Using
a Markov chain Monte Carlo routine the package then determines the
best fit theory parameter values and the corresponding statistical
error bars. The data sets we have used are the following: In terms
of CMB data, we have included in our parameter fitting the WMAP5
temperature and polarization power spectra with the routine for
computing the likelihood supplied by the WMAP
team\footnote{Available at the LAMBDA website:
http://lambda.gsfc.nasa.gov/.}. We also included some small-scale
CMB measurements, such as BOOMERanG \cite{BOOMERanG}, CBI
\cite{CBI}, VSA \cite{VSA} and the newly released ACBAR data
\cite{ACBAR}. Besides the CMB information, we also combine
information about the matter power spectrum from the ``Luminous Red
Galaxies" sample from the Sloan Digital Sky
Survey\cite{Tegmark:2006az} and supernova SNIa data. In the
calculation of the likelihood from SNIa we have marginalized over
the ``nuisance parameter" \cite{SNMethod}. The supernova data we use
is the ``Union" compilation (307 sample) \cite{Union}. We also
consider small scale information obtained from the Lyman-$\alpha$
forest power spectrum from the Sloan Digital Sky Survey (SDSS)
\cite{Lya}, however we also keep its unclear systematics in mind
\cite{WMAP5GF1,FogliWMAP5}. Furthermore, we make use of the Hubble
Space Telescope (HST) measurement of the Hubble parameter
$H_{0}\equiv 100$h~km~s$^{-1}$~Mpc$^{-1}$ by using a Gaussian
likelihood function centered around $h=0.72$ with standard deviation
$\sigma=0.08$ \cite{HST}.


\section{Numerical Results}

In this section we show the results from the global fitting to the
observational data. We focus on $k_{s}$, the scale at which the
break in the power spectrum occurs, and $n_s, n_{s2}$, the power
index parameters, and we marginalize over the other cosmological
parameters.

The numerical results are listed in Tables I. We give the results
obtained in two different calculations: in one we keep free the
parameters $n_s$ and $k_{s}$ and fix $n_{s2}=3$, in the other we
keep $n_s$, $k_{s}$ and $n_{s2}$ all to be free parameters.

\begin{table} \hspace{-5mm}
TABLE I. Constraints on the parameters $n_s$ and $k_{s}$ from CMB,
LSS, SNe with/without Lyman $\alpha$ data. We have considered two
cases: either fixing $n_{s2} = 3$ or keeping $n_{s2}$ free. The results are
obtained by marginalizing over the other theory parameters.
\begin{center} 

\begin{tabular}{|c|c|c|c|c|}

\hline

&\multicolumn{2}{|c|}{CMB$+$LSS$+$SN} & \multicolumn{2}{|c|}{CMB$+$LSS$+$SN$+$Lya }\\

\cline{2-5}

& fix $n_{s2}=3$& free $n_{s2}$&fix $n_{s2}=3$&free $n_{s2}$\\

\hline

$n_s$&$0.957^{+0.0112}_{-0.0116}$&$0.957^{+0.0115}_{-0.0116}$&$0.960\pm0.0114$&$0.959^{+0.0119}_{-0.0116}$\\

\hline

$k_s$&$>0.289 (2\sigma)$&$>0.195 (2\sigma)$&$>1.02(2\sigma)$&$>1.05(2\sigma)$\\

\hline


\hline
\end{tabular}
\end{center}
\end{table}

\begin{figure}[htbp]
\begin{center}
\includegraphics[scale=0.4]{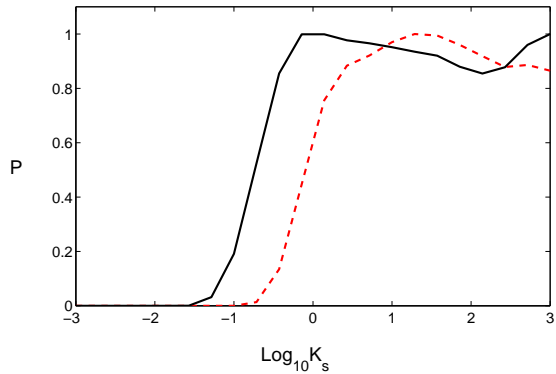}
\caption{Constraints on $k_{s}$ from the observations, assuming a
spatially flat universe. The vertical axis is the probability, the
horizontal axis co-moving scale. The black solid line shows the
probability distribution for the position $k_{st}$ of the break in
the spectrum obtained using the
CMB $+$ LSS $+$ SN Ia data. The red dashed line is obtained
by additionally taking into account the Lya data.\label{fig1}}
\end{center}
\end{figure}

In Fig. \ref{fig1} we plot the one dimensional probability
distribution of $k_{s}$ obtained by fitting with observational data. In this
calculation, we have fixed the second power index to be $n_{s2}=3$
and let the first index $n_s$ as well as $k_{s}$ be free. The
black solid line is obtained if we only use the CMB, LSS and SN Ia
data, the red dashed line results by including in addition the Lyman
$\alpha$ data. The probability distribution is obtained by
marginalizing over the other cosmological parameters.

We find that with current data we can obtain a limit on $k_{s}$.
Since the data (in particular the Lyman $\alpha$ data) tends to
indicate that there is slightly less structure on small scales than
a scale-invariant primordial power spectrum would predict, and the
matter bounce model predicts more power on small scales than is
obtained from a scale-invariant spectrum, we get a lower bound on
the k-value of the break point. From CMB, LSS and SN data alone, we
get $k_{s} > 0.289 h~Mpc^{-1}$ at $2\sigma ~ C. L.$. Taking into
account the Lyman $\alpha$ data, the limit on $k_{s}$ can be pushed
to a much smaller length scale: the $2\sigma$ limits  can be pushed
to $k_{s} > 1.02 h~Mpc^{-1}$.

\begin{figure}[htbp]
\begin{center}
\includegraphics[scale=0.4]{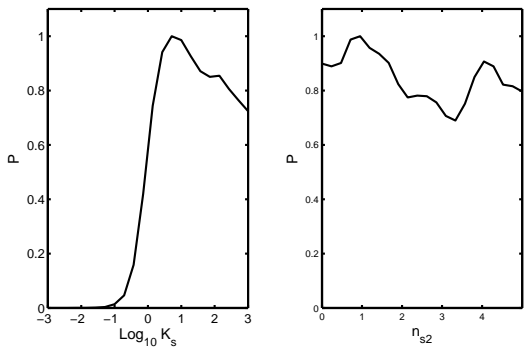}
\caption{Constraints on $k_{s}$ and the second power index $n_{s2}$
from  CMB $+$ LSS $+$ SN Ia $+$ Lya data. \label{fig2}}
\end{center}
\end{figure}

In Fig. \ref{fig2} we plot the results obtained by keeping both
$n_{s2}$ and $k_{s}$ free. We find there are only weak constraints
on $n_{s2}$. The data mildly prefers a red rather than a blue tilt
(such as the ``matter bounce" would predict), and prefers the
position of the break in the power spectrum to be at the value $k_s
= 19~h~Mpc^{-1}$. The interpretation of this result is that a
primordial spectrum which is almost scale-invariant, characterized
by a fixed power index $n_s$, fits the large-scale structure data
well, but that there is less structure on scales below the break
point.

\section{Summary}
\label{Sum}

There are theoretical models such as the ``matter bounce" which
predict a transition in the power spectrum from being approximately
scale-invariant on length scales larger than some distinguished
scale to being approximately scaling as $n_s = 3$ in shorter length
scales. In this note, we have studied the constraints on the value
of $k_s$, the co-moving momentum at which the break occurs. The
current cosmological data put a $2\sigma$ upper limit which is $k_s
\sim o(1)~h Mpc^{-1}$. Future data sets will be able to set much
tighter constraints on $k_s$.


\section*{Acknowledgements}

We acknowledge the use of the Legacy Archive for Microwave
Background Data Analysis (LAMBDA). Support for LAMBDA is provided by
the NASA Office of Space Science. We have performed our numerical
analysis on the Shanghai Supercomputer Center (SSC). We thank Yi-Fu
Cai for helpful discussions. This work is supported in part by the
National Natural Science Foundation of China under Grant Nos.
90303004, 10533010 and 10675136 and by the Chinese Academy of
Science under Grant No. KJCX3-SYW-N2. RB wishes to thank the
Institute of High Energy Physics for hospitality and financial support,
and also the KITPC for hospitality and support during the program
``Connecting Fundamental Physics with Cosmological Observations".
RB is also supported by an NSERC Discovery Grant and by funds
from the CRC program.


\end{document}